\documentclass[%
 reprint,
 amsmath,amssymb,
 aps,
 prm,
]{revtex4-2}

\usepackage{graphicx}
\usepackage{dcolumn}
\usepackage{bm}

\begin{document}

\preprint{}

\title{Enhanced Superconductivity in the Se-substituted 1T-PdTe$_2$ }

\author{Wenhao Liu$^1$}
\author{Sheng Li$^1$}
\author{Hanlin Wu$^1$}
\author{Nikhil Dhale$^1$}
\author{Pawan Koirala$^1$}
\author{Bing Lv$^{1,2 }$}%
\email{blv@utdallas.edu}
\affiliation{%
$^1$Department of Physics, University of Texas at Dallas, Richardson, Texas 75080, USA\\
$^2$Department of Materials Science and Engineering, University of Texas at Dallas, Richardson, Texas 75080, USA
}%

\date{}

\begin{abstract}
  Two-dimensional transition metal dichalcogenide PdTe$_2$ recently attracts
 much attention due to its phase coexistence of type-II Dirac semimetal and type-I superconductivity. 
 Here we report a 67 \% enhancement of superconducting transition temperature in the 1T-PdSeTe in comparison to that of PdTe$_2$ through partial substitution of Te atoms by Se.
 The superconductivity has been unambiguously confirmed by the magnetization, resistivity and specific heat measurements. 
 1T-PdSeTe shows type-II superconductivity with large anisotropy and non-bulk superconductivity nature with volume fraction $\approx$ 20 \% estimated from magnetic and heat capacity measurements. 
 1T-PdSeTe expands the family of superconducting transition metal dichalcogenides and thus provides additional insights for understanding superconductivity and topological physics in the 1T-PdTe$_2$ system

\end{abstract}

\maketitle


\section{\label{sec:level1}INTRODUCTION}

Recently, two-dimensional (2D) materials, including transition metal dichalcogenides (TMDs) have received considerable attention due to their emergent physical properties, which in turn have potential to revolutionize many fields in both fundamental science and technological applications \cite{S1, S2, S3, S4}. For example, a large current on/off ratio exceeding an order of 10$^8$ and ultrahigh mobility are found in MoS$_2$ \cite{S5}; Large transverse magnetoresistance induced by chiral anomaly was observed in MoTe$_2$ and WTe$_2$ \cite{S6, S7}; and  2D magnetism down to atomic layers exists such as VSe$_2$ \cite{S8, S9}. The studies of superconductivity through chemical intercalation/doping and the interplay with different sets of CDW orders have stimulated many research interests ever since the 1970s \cite{S10, S11, S12, S13, S14, S15}. Recently, such studies have been extended to electrostatic gating where both the \textit{T$_c$} and CDW transition could be significantly tuned through the gating voltage \cite{S16, S17, S18, S19}.Further studies on the TMDs at atomically thin layer limit result in the Ising pairing in NbSe$_2$ and Wigner crystal, minibands or other exotic states in the twisted angles TMD heterostructures \cite{S20, S21, S22, S23}.

Structurally, TMDs are constructed by either octahedral or trigonal prismatic MX$_2$ layers where M is a transition metal from groups 4 to 10 elements and X is a chalcogen. There is strong X-M-X bonding within each layer, while a rather weak van der Waals interactions between different layers, which result in the easily exfoliable nature for most TMD materials. The variations of stacking sequences of MX$_2$ layers as well as the atomic coordination often lead to different polytype and polymorphic structures. Based on the numbers of layers in the unit cell and their symmetry, the most common polytypes are 1T (T: trigonal), 1T$'$ (distorted 1T), 2H (H: hexagonal) and 3R (R: rhombohedral) respectively.  Among these different TMDs, PdTe$_2$ has attracted much attention lately as it has been experimentally verified as type-II Dirac semimetal \cite{S24, S25}.

Dirac semimetals show a Dirac cone where a valence band and a conducting band touch each other at one point (Dirac point) in the energy-momentum space. The low energy quasiparticle excitations near the Dirac point is called a Dirac Fermion, which can be well described by a relativistic Dirac Equation. Remarkably, superconductivity is also discovered in PdTe$_2$ system \cite{S26}. Specific heat and dc magnetization measurements reveal that PdTe$_2$ is a conventional type-I superconductor with a transition temperature $\sim$1.64 K \cite{S27, S28}. Besides, a new type Ising superconductivity is found in ambient-stable PdTe$_2$ film with in-plane centrosymmetry, showing large in-plane critical field more than 7 times the Pauli limit \cite{S29}. The simultaneously robust superconductivity as well as the nontrivial topological states make PdTe$_2$ a great potential candidate in electronic applications, such as the realization of topological superconductors, which can be used in fault-tolerant Quantum Computing and potentially lead to important technological applications. 
In this paper, we reported our research effort to further enhance the superconductivity in the topological PdTe$_2$ system through chemical doping, and we found significantly enhanced superconductivity in the Se-doped 1T-PdSeTe for the first time. The structure was confirmed by powder X-ray diffraction, which suggests 1T-PdSeTe are crystallized in the same CdI$_2$ type structure with space group P$\bar{3}$m1\ (164) as its PdTe$_2$ counterpart. Magnetization revealed large anisotropic superconducting shielding fractions along different magnetic orientations. Specific heat measurements along with magnetization furtherly demonstrated non-bulk feathers of the superconductivity in 1T-PdSeTe. The superconducting phase transition temperature (\textit{T$_c$}), determined from the resistivity results is 2.74 K, which is about 67$\%$ enhancements compared to that of PdTe$_2$. The discovery of superconducting 1T-PdSeTe expands the superconducting TMD family and could provide another planform to study the non-trivial topological superconductivity in the TMDs.

\section{\label{sec:level2}EXPERIMENT DETAILS}

1T-PdSeTe is firstly reported in 1965 \cite{hulliger}. Our 1T-PdSeTe specimens were synthesized by a self-flux method through melting stoichiometric amount of Pd (99.95\%, Alfa Aesar), Se (99.999\%, Alfa Aesar) and Te (99.999\%, Alfa Aesar) elements. Small Pd ingots, Se shots and Te pieces in stoichiometric ratio were weighted inside Ar glovebox with a total moisture and oxygen level less than 0.1 ppm. Se shots and Te pieces were ground into fine powders and loaded inside the silica tube together with Pd ingots. The silica tube was flame-sealed under vacuum and placed in a single-zone furnace. The assembly was slowly heated up to 800$^\circ$C, held for three days,  then followed by furnace cooling to room temperature. Sizeable shiny and platelet-like crystals (up to 3 mm size) with preferred crystallographic \textit{c} axis orientation are obtained by cleaving ingot product. 

Powder X-ray diffraction (XRD) measurements were performed on crushed crystals using a Rigaku SmartLab X-ray diffractometer equipped with Cu-K$\alpha$ radiation at room temperature and Rietveld refinement was carried out using GSAS-II \cite{S30}. Scanning Electron Microscopy (SEM) with Energy Dispersive X-ray Analysis (EDX) is performed using Zeiss-LEO 1530. Measurement of direct -current (DC) magnetization was conducted by Quantum Design Magnetic Property Measurement System (MPMS) down to 1.8 K. Four gold wires (50 $\mu$m, in diameter) were pasted on the freshly cleaved sample surface by silver epoxy as four probes. Resistivity was measured with a four-probe method using Quantum Design Physical Property Measurement System (PPMS) down to 1.8 K. Specific heat measurement was done in the PPMS by a time-relaxation method with magnetic field parallel to the crystallographic \emph{c} axis.

\section{\label{sec:level3}RESULTS}

\begin{figure}[b]
\includegraphics[width=8.5cm]{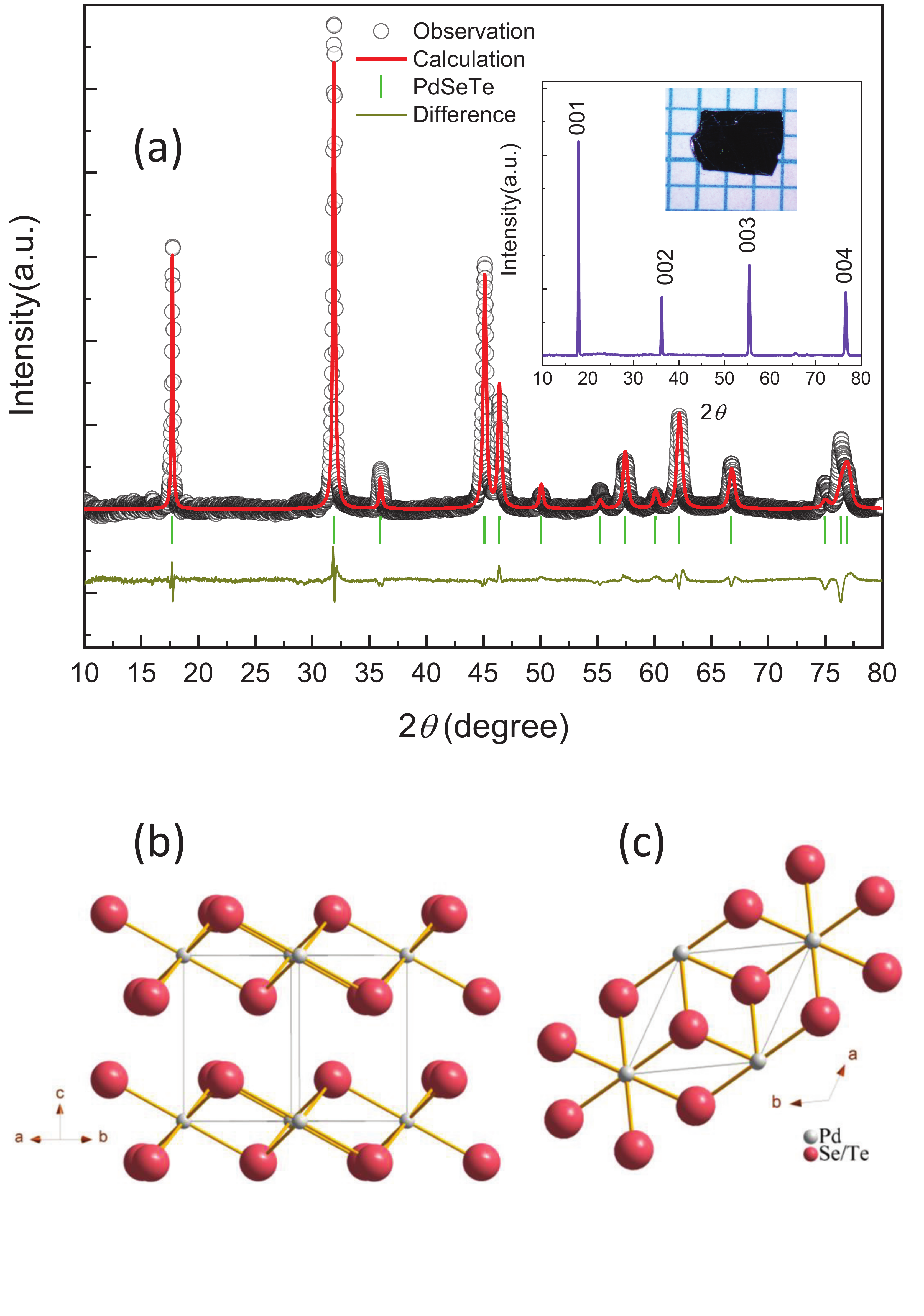}
\caption{\label{fig1} (a) XRD pattern and Rietveld refinement of 1T-PdSeTe at room temperature. Vertical marks (green bars) stand for positions of Bragg peaks. The inset shows XRD pattern and optical image on the mm size grid of the easily exfoliated flake. Ball and stick model of the crystal structure of 1T-PdSeTe showing the side view (b) and top view (c).}
\end{figure}

The X-ray diffraction patterns on different crystals from the same batch are nearly identical and show only (00l) peaks, i.e., crystallographic \textit{c} axis is preferred orientation, as shown in Fig.~\ref{fig1} inset. This indicates that \textit{c} direction is perpendicular to the flat surface of platelet-shaped crystals. The full Rietveld refinement of powder X-ray diffraction on crushed powders from the single crystals is shown in Fig.\ref{fig1}, which establishes that PdSeTe indeed adopts the same crystal structure as 1T-PdTe$_2$ in trigonal space group of \emph{P}$\bar{3}$\emph{m}1(164). Good refinement values of \textit{R$_p$} = 1.91\%, and  \textit{R$_{wp}$} = 3.72\% indicate a good agreement of experimental data with the refined structural model and high quality of our samples. Detailed refinement results with associated Wyckoff position of each atom are shown in Table~\ref{tab1}. The refined lattice parameters are \textit{a} = \textit{b} = 3.9029(8) \AA, \textit{c} = 4.980(1) \AA, $\alpha$ = $\beta$ = 90$^\circ$, $\gamma$ = 120$^\circ$  for 1T-PdSeTe. These values are a bit smaller than those of isostructural PdTe$_2$ (\textit{a}=\textit{b}=4.03 \AA, \textit{c}=5.12 \AA) as expected, since Te atoms are replaced by smaller Se atoms in the unit cell. Refined occupancies of Se and Te are 0.47(2) and 0.53(2) respectively, and very close to the nominal Se: Te ratio. Fig.~\ref{fig1} (b) and (c) show ball and stick models of crystal structure of 1T-PdSeTe from the refinement results. Pd atom is found in octahedral coordination with Se/Te atoms in each layer and every layer stacks repeatedly via vdW forces along \textit{c} axis of the trigonal unit cell. SEM analysis of crystals on multiple points shows only existence of Pd, Se and Te elements in the crystals, and energy-dispersive X-ray spectroscopy (EDS) shows a ratio of Pd: Se: Te=1: 0.92: 1, consistent with Rietveld refinement results and suggests slight Se vacancies might exist in our samples.

\begin{table}[b]
\renewcommand\arraystretch{1.5}
\caption{\label{tab1}
Summary of Crystallographic data and structure refinement parameters for 1T-PdSeTe.}
\begin{ruledtabular}

\begin{tabular}{cccccccc}

&Compounds &  & 1T-PdSeTe\\
\hline

&Space group & & \emph{P}$\bar{3}$\emph{m}1\\
& \textit{a} (\AA)  & & 3.9029(8) \\
& \textit{b} (\AA) & & 3.9029(8)\\
& \textit{c} (\AA) & & 4.980(1)\\
& \textit{V} (\AA$^3$) &  &65.699(4)\\
& \textit{D$_{cal}$} (g/cm$^{3} $)& &7.94 \\
& \textit{R}$_{p}$ & &1.91 \% \\
& \textit{R$_{wp}$} & &3.72 \% \\
\\

 &Atom&Site&Symm.&x/a&y/b&z/c&Occ.\\
\hline
&Pd1 & 1a & $\bar{3}$m & 0 & 0 & 0 & 1 \\
&Se1 & 2d & 3m & 1/3 & 1/3 & 0.2537(3) & 0.47(2)\\
&Te1 & 2d & 3m & 1/3 & 1/3 & 0.2537(3) &0.53(2)\\
\end{tabular}
\end{ruledtabular}
\end{table}

\begin{figure}[b]
\includegraphics[width=8.5cm]{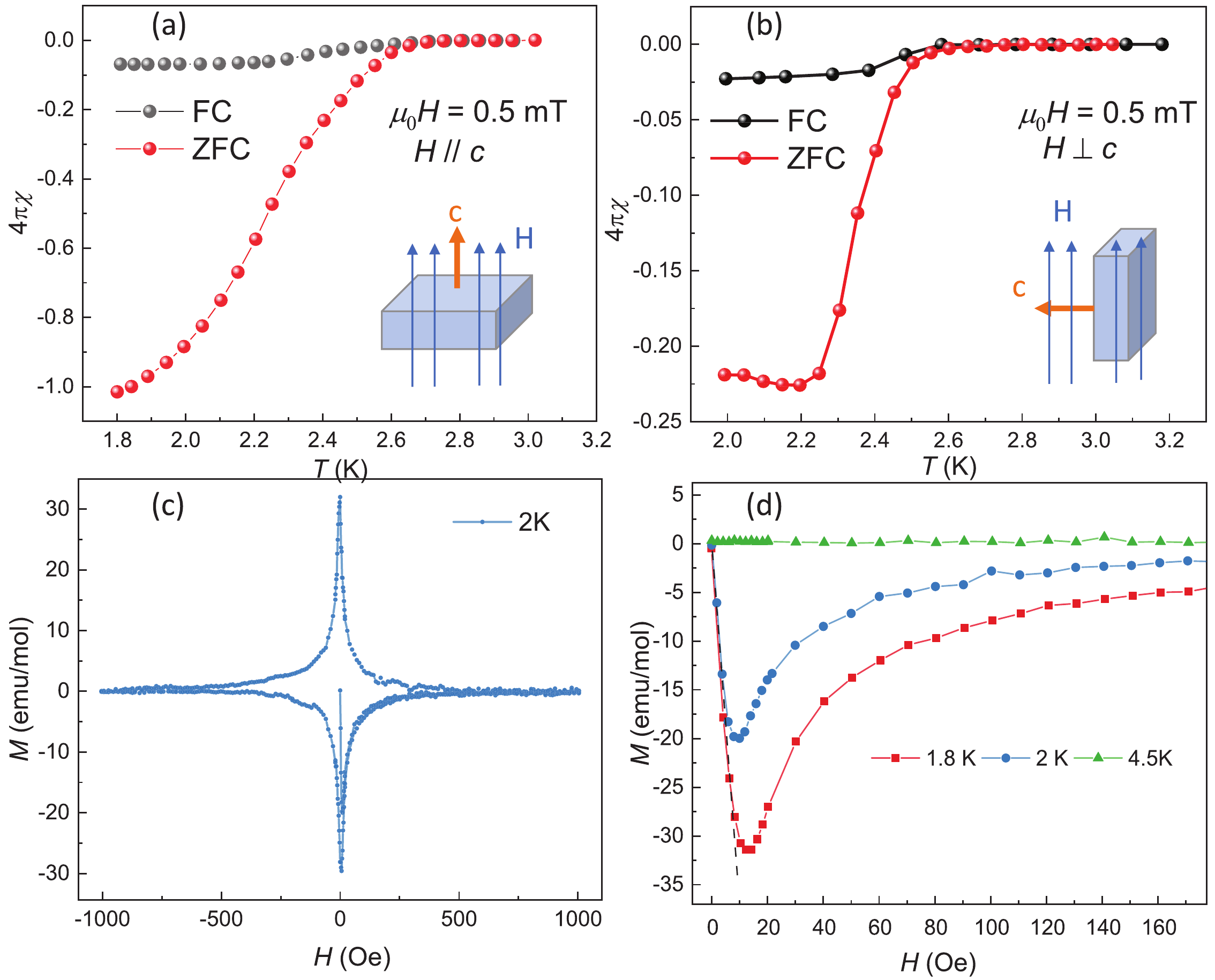}
\caption{\label{fig2}Temperature dependence of magnetic susceptibility measured in both ZFC mode and FC mode with an external field of 5 Oe parallel to the crystallographic \emph{c} axis (a) and perpendicular to the \emph{c} axis (b). (c) Magnetic hysteresis loop of 1T-PdSeTe at 2 K with external field parallel to the \emph{c} axis. (d) MHLs in the magnetic penetration process at 1.8 K, 2 K and 4.5 K with external field parallel to \emph{c} axis. The dashed line represents the slope of the perfect diamagnetism.}
\end{figure}

Fig.~\ref{fig2}(a) and (b) illustrate temperature-dependent magnetic susceptibility measured in zero-field-cooled (ZFC) and field-cooled (FC) modes under an external field of 5 Oe parallel to crystallographic \emph{c} axis and perpendicular to \emph{c} axis respectively. A clear superconducting transition can be seen in both modes. When magnetic field is parallel to \emph{c} axis with the large demagnetization enhancement factor, a large shielding volume fraction $\sim$ 102 \% at 1.8 K is observed using an estimated density 7.92 g/cm$^3$. However, when the magnetic field is parallel to the \emph{a-b} plane where demagnetization enhancement factor is minimized, the actual shielding fraction we obtained is only about 20 \% for the same crystal, reflecting real superconducting volume fraction is less than 100 \% and non-bulk superconductivity nature exists in this system. Further annealing the crystals at 500 $^\circ$C for a few days does not improve the superconducting volume fraction. In Fig.~\ref{fig2}(c), we show the magnetization hysteresis loop (MHL) of 1T-PdSeTe at 2 K with a magnetic field applied parallel to \emph{c} axis. Fig.~\ref{fig2}(d) shows MHLs in the magnetic penetration process at 1.8 K, 2 K and 4.5 K with a magnetic field applied parallel to the \emph{c} axis. The shielding behavior is suppressed with increasing temperature and disappears when the temperature is higher than \emph{T$_c$}. Fully penetrating fields corresponding to the maximum value of magnetization, is only about 10 Oe at 1.8 K. This suggests either an easy vortex motion with weak flux pinning effect or a low charge carrier density in this system. The whole MHL indicates that 1T-PdSeTe is a type-II superconductor, in strong contrast with the type-I superconductivity found in isostructural 1T-PdTe$_2$ compound \cite{S27}. In addition, the slopes for the perfect diamagnetism at 1.8 K and 2K are quite close to each other. The superconducting volume fraction estimated from the M-H curve is about 111\%, which is quite close to the 103\% estimated from the M-T results.

\begin{figure}[b]
\includegraphics[width=8.5cm]{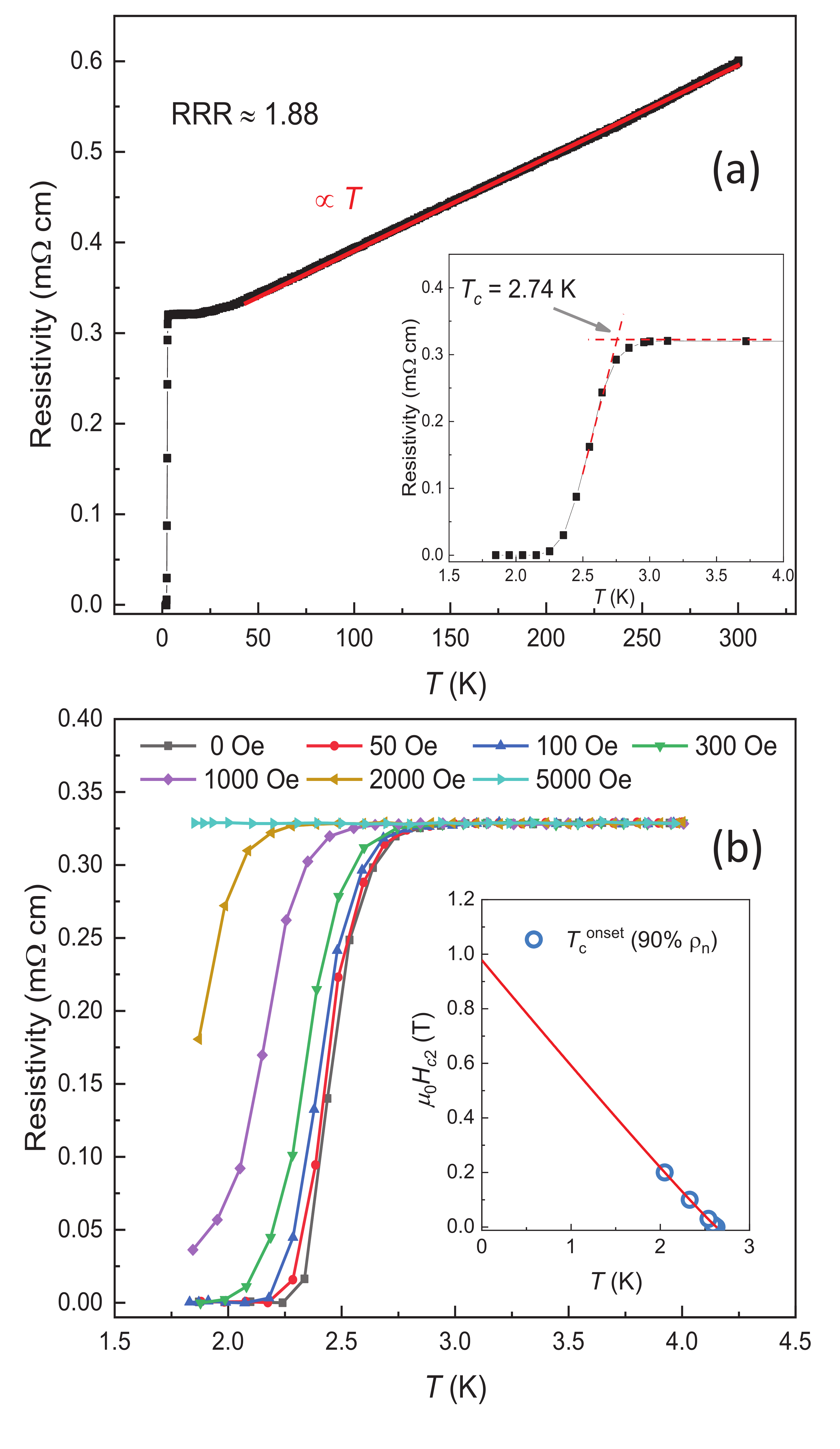}
\caption{\label{fig3}(a) Temperature dependence of resistivity for 1T-PdSeTe. The inset shows an enlarged view of the resistivity near the superconducting transition temperature. (b) Temperature dependence of resistivity for 1T-PdSeTe at zero field and various magnetic fields parallel to \emph{c} axis. The inset shows upper critical field $\emph{H}_{\emph{c}2}$ as a function of temperature. The red line represents the best fitting using the function described in the text. }
\end{figure}

Temperature dependent in-plane resistivity ($\rho_{ab}$) of 1T-PdSeTe is shown in Fig.~\ref{fig3}(a). Resistivity decreases nearly linearly with temperature down to 50 K and keeps almost the same value when temperature continues to decrease till \emph{T$_c$}. The Residual Resistivity Ratio (\emph{RRR}) defined as \emph{RRR} = \emph{R}$_{300\emph{K}}$/\emph{R}$_{4\emph{K}}$ is 1.88 for our 1T-PdSeTe samples. The \emph{RRR} is smaller than that of pure PdTe$_2$ (\emph{RRR} $\approx$ 30)\cite{S31}, which is possibly due to the increase of electron scattering by lattice distortions or defects induced by Se doping. In the inset of Fig.~\ref{fig3}(a), an enlarged view of resistivity near the transition temperature is presented, clearly showing the superconducting transition at about 2.74 K, which is a distinct 67\% enhancement of \emph{T$_c$} from that of PdTe$_2$ (1.64 K). Fig.~\ref{fig3}(b) shows magnetic field dependence of resistivity for 1T-PdSeTe. \emph{T$_c$} is gradually suppressed when the external field increases, a classical characteristic for superconductors. Taking 10\% of resisitivty drop as criteria, temperature dependence of the upper critical field $\emph{H}_{\emph{c}2}$ is shown in the inset of Fig.~\ref{fig3}(b). The upper critical field of 1T-PdSeTe shows a bit positive curvature close to \emph{{T$_c$}}, which is a characteristic of two-band clean-limit type-II superconductors and commonly existed in many layered superconductors, including MgB$_2$\cite{MgB2},  RNi$_2$B$_2$C\cite{prl-RNiBC, phyc-RNiBC} (R: rare earth elements), and various superconducting TMDs. Consequently, the traditional Werthamer–Helfand– Hohenberg (WHH) model, which normally has negative curvature close to \emph{T$_c$}, does not fit the data very well. Experimentally, our $\emph{H}_{\emph{c}2}$ can be fitted better in the entire \emph{T/T$_c$} range using a semi-empirical expression \emph{H}$_{\emph{c}2}$(\emph{T}) = \emph{H}$_{\emph{c}2}^\ast$(1-\emph{T/T$_c$)}$^{1+\alpha}$, previously applied to MgB$_2$, RNi$_2$B$_2$C, WTe$_2$, MoTe$_2$ and so on\cite{S12, MgB2,prl-RNiBC,phyc-RNiBC, WTe2}.  Noted that the fitting parameters \emph{B}$_{\emph{c}2}$* = 0.97(2) T can be considered as the upper limit for the upper critical field \emph{B}$_{\emph{c}2}$(0). Thus the Ginzburg-Landau coherence length $\xi_{GL}$(0) can be estimated as $\sim$ 18 nm.  

\begin{figure}[b]
\includegraphics[width=8.5cm]{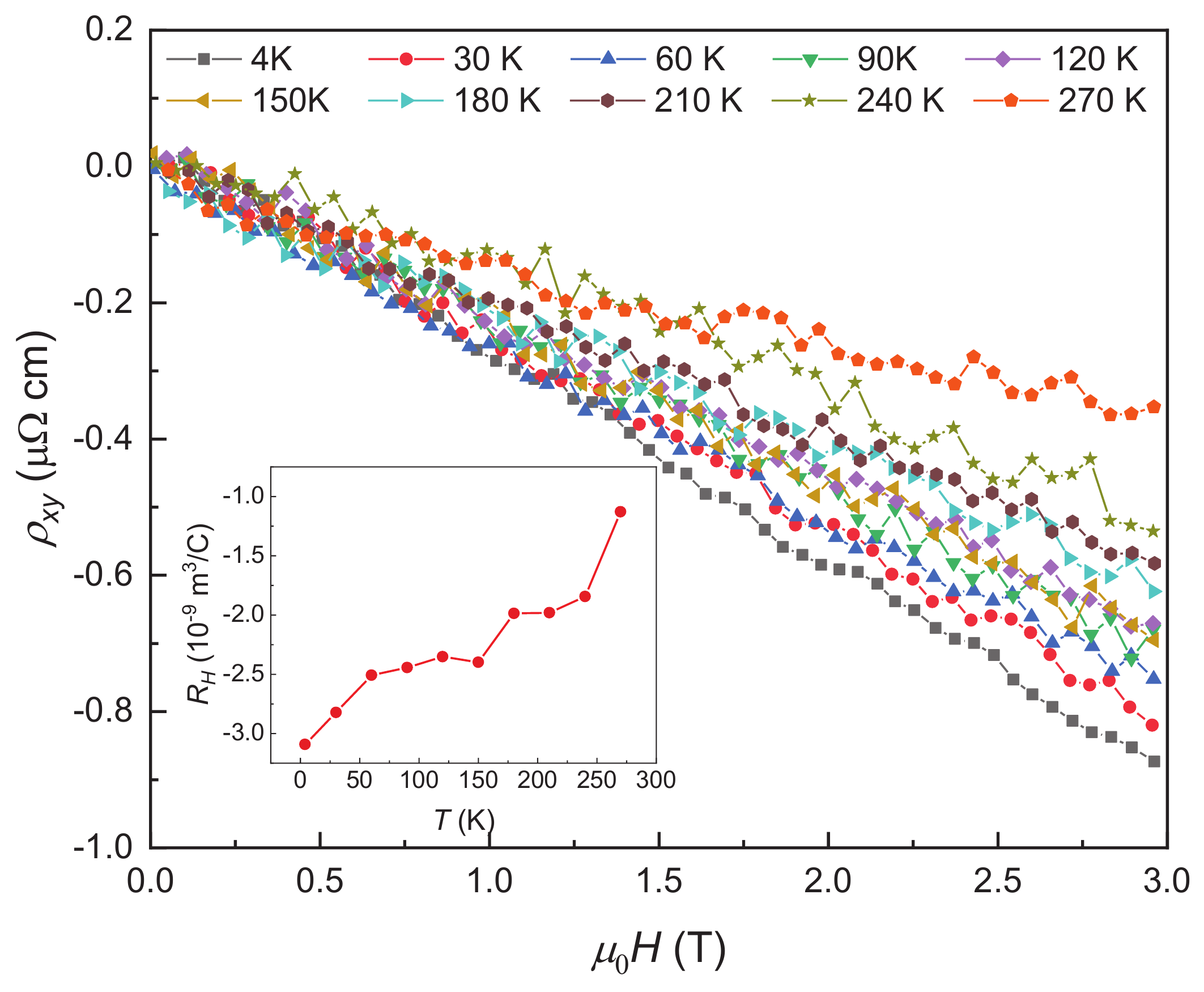}
\caption{\label{fig4} Magnetic field dependence of  Hall resistivity\emph{$\rho_{xy}$} for 1T-PdSeTe at various temperatures. The inset shows the temperature dependence of Hall coefficient \emph{$R_H$}.}
\end{figure}

Fig.~\ref{fig4} shows temperature dependence of the Hall resistivity \emph{$\rho_{xy}$} at 4, 30, 60, 90, 120, 150, 180, 210, 240 and 270 K, respectively at magnetic field up to 3 T. Here \emph{$\rho_{xy}$} was measured with a longitudinal current. The magnetic field is perpendicular to the current and plate surface of 1T-PdSeTe samples. The voltage \emph{V$_{xy}$} was measured in the direction across the sample width. As we can see, \emph{$\rho_{xy}$} shows almost linearly dependent behavior with the magnetic field. The slope is all negative from 4 to 270 K, indicating that charge carriers are dominantly electrons at the Fermi surface. We estimated Hall coefficient \emph{$R_H$}  = \emph{$\rho_{xy}$} /$\mu_0$\emph{H} by linearly fitting the \emph{$\rho_{xy}$} at various temperatures, as is shown in the inset of Fig.~\ref{fig4}. Besides, the \emph{$R_H$} scales with the temperature almost monotonically and such monotonic dependence behaviors is consistent with many other TMD materials irrespective of the positive or negative sign of \emph{$R_H$}\cite{S18}. The charge-carrier density estimated from \emph{n} = 1/(\emph{e} $\times $\emph{R$_H$})  is about 6.25 $\times$ 10$^{21}$ cm$^{-3}$ at 270 K and 1.95 $\times$ 10$^{21}$ cm$^{-3}$ at 4 K, respectively. Compared to the reported value of about 5.5 $\times$ 10$^{21}$ cm$^{-3}$ of PdTe$_2$ \cite{S27}, the carrier density does not change too much as Se does not introduce extra carriers when doped at Te sites in PdTe$_2$. This further suggests that rather than the charge carriers, the admirable \emph{{T$_c$}} enhancement might originate from other factors, which we will discuss further later.  

\begin{figure}[b]
\includegraphics[width=8.5cm]{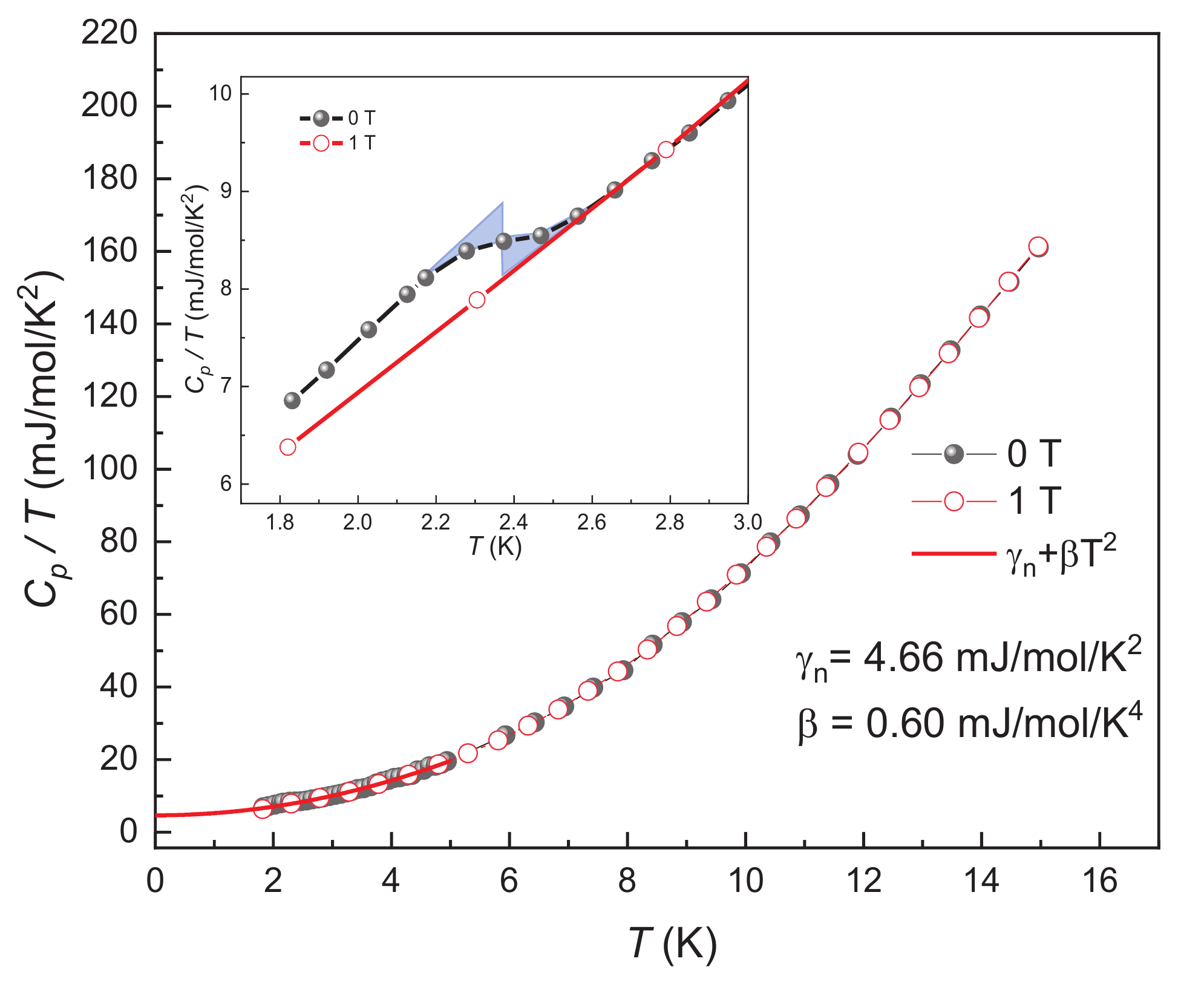}
\caption{\label{fig5}(a) Temperature dependence of specific heat under 0 T (solid circles) and 1 T (open circles). The red line represents the fitting using \emph{C$_p$/T} = \emph{$\gamma_n$}+\emph{$\beta$T}$^2$. The inset shows an enlarged view of the temperature dependence of specific heat below 3 K. The \emph{T$_c$} is obtained utilizing the equal area construction method for specific heat data.}
\end{figure}

In order to get more insights about electronic and superconducting properties of our sample, we have carried out the specific heat measurement. Fig.~\ref{fig5} shows raw data of specific heat of 1T-PdSeTe at 0 T and 1 T. Only a small specific heat anomaly appears in low temperature region starting from 2.6 K, as shown in the inset of Fig.~\ref{fig5}. This anomaly is coincident with the superconducting transition probed by resistivity and magnetization measurements. At 1 T, the superconductivity is completely suppressed, as suggested by the linear \emph{C$_p$/T} vs \emph{T} plot in the inset of Fig.~\ref{fig5}. For 1T-PdSeTe, the normal state data can be fitted quite well using the Debye model \emph{C$_p$/T} = \emph{$\gamma_n$}+\emph{$\beta$T}$^2$, which yields $\gamma_n$ = 4.66\ mJ/mol/K$^2$, $\beta$ = 0.60 mJ/mol/K$^4$. The fitting curve is shown as the red line in Fig.~\ref{fig5}. Debye temperature $\Theta_D$ can be calculated from the $\beta$ value through the relationship $\Theta_D$=(12$\pi$$^4N_Ak_BZ$/(5$\beta$))$^{1/3}$, where N$_A$ is the Avogadro constant, and Z is the number of atoms in one unit cell, which is 3 for 1T-PdSeTe. The obtained Debye temperature of 1T-PdSeTe is about 213 K. It is worthwhile to note that the residual carrier contribution is estimated as $\gamma_0$ = 3.71 mJ/mol/K$^2$ and superconducting volume fraction therefore estimated from specific heat data is about ($\gamma_n-\gamma_0)$/$\gamma_n$ $\approx$ 20 \%, which is quite consistent with the magnetic susceptibility measurement when \emph{H // ab} and further support non-bulk superconductivity in this compound. 

Besides, derived Debye temperature for 1T-PdSeTe is slightly higher than that of PdTe$_2$ ($\Theta_D$ = 207 K) single crystals\cite{S28}. The average electron-phonon coupling can be roughly estimated from the McMillan equation:
\begin{eqnarray}
\lambda_{ep}=\frac{\mu^\ast I n\left(\frac{1.45\ T_c}{\Theta_D}\right)-1.04}{1.04+In\left(\frac{{1.45\ T}_c}{\Theta_D}\right)\left(1-0.62\mu^\ast\right)}
\label{eq:one}.
\end{eqnarray}

Similar to other TMD material 2H-TaSe$_{2-x}$S$_x$\cite{S13}, we assume the empirical value of the Coulomb pseudopotential $\mu^\ast$ = 0.15. Utilizing this Coulomb pseudopotential and the calculated Debye temperature, we can estimate $\lambda_{ep}$ of 0.636 and 0.572 for 1T-PdSeTe and PdTe$_2$ respectively. 

\section{\label{sec:level4}DISCUSSION}

\begin{figure}[b]
\includegraphics[width=8.5cm]{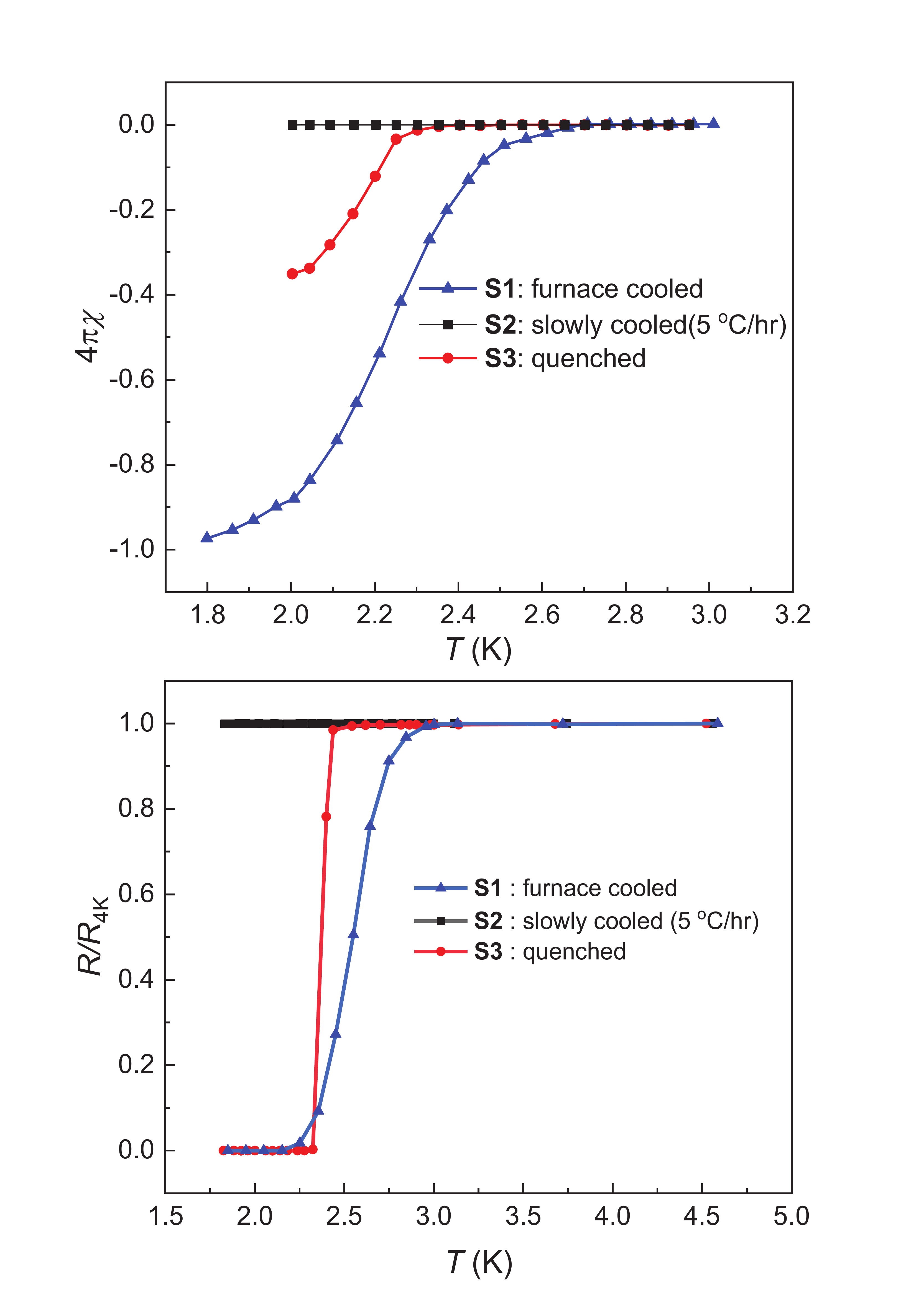}
\caption{\label{fig6}(a) Temperature dependence of magnetization for 1T-PdSeTe near superconducting phase transition with external magnetic field parallel to the \emph{c} axis. The inset shows an enlarged view of the temperature dependent magnetization of \textbf{S2} and \textbf{S3} in ZFC mode. (b) Temperature dependence of normalized resistivity for 1T-PdSeTe synthesized under different cooling conditions.}
\end{figure}

The significantly enhanced \emph{T$_c$} and non-bulk superconductivity in our 1T-PdSeTe sample is rather intriguing. Conventionally, the \emph{T$_c$} enhancement is mainly caused by increasing charge carriers through chemical doping or physical pressure. However, our Hall measurement suggests that Se substitution does not introduce extra charge carriers. Although smaller Se substitution into PdTe$_2$ lattice could cause the chemical pressure that plays a similar role as external pressure to tune the electronic structure, it is reported that transition temperature of PdTe$_2$ can only be enhanced up to 1.91 K under an external pressure around 0.91 GPa. The \emph{T$_c$} in fact gradually decreases to 1.27 K when the external pressure reaches 2.5 GPa \cite{S31}. This further suggests that chemical pressure caused by Se substitution might not be sufficient to enhance \emph{T$_c$} up to 2.74 K in 1T-PdSeTe phase. On the other hand, the slightly larger $\lambda_{ep}$ of 1T-PdSeTe (i.e., stronger electron-phonon coupling) than that of 1T-PdTe$_2$, might be supportive of \emph{T$_c$} enhancement but cannot be the root cause for the enhanced superconductivity. Another possibility for this enhanced superconductivity could be lattice disorder and/or structural defects caused by Se doping. This phenomena have been observed recently in many low dimensional materials including TMDs. For example, superconducting transition temperature of TaS$_2$ can be enhanced from 2.89 K to 3.61 K due to the disorder arising from the structural defects \cite{S34}. Bulk superconductivity with enhanced \emph{T$_c$} was achieved in ZrTe$_3$ through growth induced structural disorder \cite{S35}. To further elucidate such possibilities, we have carried out some control experiments as follows: three batches of samples with the same amount of reactants, the same heat treatments but three different cooling rates, are presented. Specimen1 (\textbf{S1}) samples are cooled in furnace by turning off the power at 800 $^\circ$C as shown before; Specimen 2 (\textbf{S2}) undergo slowly cooling with a rate of  5 $^\circ$C/hour from 800 $^\circ$C down to room temperature which presumably should have the least grain boundaries and point defects, and Specimen 3 (\textbf{S3}) samples are quenched into ice water batch directly at 800 $^\circ$C, where the presence of structural disorder and defects should be the most among all three specimens. Surprisingly, both \textbf{S2} and \textbf{S3} show worse superconductivity signals compared to \textbf{S1}. As shown in Fig.~\ref{fig6}(a), under the same measurement condition where an external magnetic field is parallel to the \emph{c} axis, no diamagnetic shielding (i.e., no superconductivity) is observed in the crystals from \textbf{S2}. And \textbf{S3} is superconducting with a much small volume fraction ($\approx$ 35\%) compared to that of \textbf{S1}($\sim$100\%). This suggests that disorder/defects need to reach a certain threshold for the emergence of superconductivity with enhanced \emph{T$_c$} , but certainly too much disorder will be harmful for the superconductivity. On the other hand, from the resistivity measurement, it is quite interesting that \textbf{S3} with a lower onset \emph{T$_c$}, displays a much sharper superconducting transition than that of \textbf{S1} crystal. This is reminiscent of the behavior of Ru-riched Sr$_2$RuO$_4$ superconductors. The \emph{T$_c$} of Sr$_2$RuO$_4$ was raised from 1.5 K to 3 K in Ru-rich samples with a broader transition width, where excess Ru comes out from the pure Sr$_2$RuO$_4$ and forms lamellar structures, typical of a eutectic solidification \cite{Sr2RuO4prl}. Coincidently, the 3-K Ru-rich phase is normally obtained through a faster growth rate than the 1.5-K phase for pure Sr$_2$RuO$_4$ samples. Further chemical analysis suggests that the three specimens have a slight difference in the Se content in spite of the nearly identical Pd and Te composition. The ratio of Se content is \textbf{S1}: \textbf{S2}: \textbf{S3} = 0.92(2): 0.99(6): 0.84(3). The \textbf{S1} with the highest \emph{T$_c$} and volume fraction has slightly Se deficiency, but apparently a higher Se deficient specimen \textbf{S3} does not have better superconducting signals. Therefore, it is reasonable to speculate on several scenarios or maybe their combined effects that are responsible for the enhanced \emph{T$_c$} in our furnace-cooled 1T-PdSeTe samples: 1) possible existence of metallic Pd or PdTe stripes similar to the Ru-Sr$_2$RuO$_4$ \cite{Sr2RuO4prl,S41,S42}; 2) some dedicated structural disorder caused by a rather small amount of Se deficiency \cite{S35,S43,S44}. 
The exact origin and mechanism could not be pinned down at this point unfortunately, which is subject to future studies due to our current research constraints. It is also worthwhile to investigate whether the type-II Dirac band dispersion, as seen in the 1T-PdTe$_2$ phase, is preserved in this 1T-PdSeTe phase in the future through band structure calculation and advanced spectroscopy studies.

\section{\label{sec:level5}CONCLUSION}
In conclusion, we have demonstrated that the superconducting transition temperature in the isostructural 1T-PdSeTe can be significantly enhanced to 2.74 K compared to the parent phase type-II Dirac semimetal PdTe$_2$ (1.64 K). The superconductivity has been investigated through resistivity, magnetization and specific heat measurement. Non-bulk superconductivity with volume fraction $\sim$20\% is demonstrated by both magnetization and specific heat data. The possibilities of the substantial \emph{T$_c$} enhancement arising from the possible existence of metallic Pd or PdTe stripes similar to the Ru-Sr$_2$RuO$_4$ and/or some dedicated structural disorder caused by a rather small amount of Se deficiency, have been examined and discussed.

\begin{acknowledgments}
This work at University of Texas at Dallas is supported by US Air Force Office of Scientific Research Grant No. FA9550-19-1-0037 and National Science Foundation (DMR-1921581). We also acknowledge the support from Office of Research at University of Texas at Dallas through Seed Program for Interdisciplinary Research (SPIRe) and the Core Facility Voucher Program. 
\end{acknowledgments}

\end{document}